\documentclass[a4paper,11pt]{scrartcl}
\usepackage{ILD}
\usepackage{tagging}
\usetag{ILD}
\usepackage{subcaption}
\usepackage{hyperref}
\usepackage{wasysym}
\usepackage{slashed}
\usepackage{wrapfig,rotating}
\usepackage{pgfpages}
\usepackage{fancybox,graphicx}
\usepackage{graphbox}
\usepackage{epstopdf}
\AppendGraphicsExtensions{.eps.gz}
\title{
   New physics searches with the International Large Detector at the ILC

}
\ildproc{phys}{2021}{012}
\date{\today}

\addauthor{
  Mikael Berggren

}
{\institute{1}}

\addinstitute{1}{
    Deutsches Elektronen-Synchrotron DESY,\\
  Notkestr. 85, 22607 Hamburg, Germany

}
\onbehalfof{
    \tagged{POS}{On behalf of }
the ILD concept group.
\tagged{POS}{The support of the LCC generator group, the ILD
software working group, the EGI federation and the Open Science GRID (via
the ILC virtual organisation) is acknowledged.}

}

\abstract{
  Although the LHC experiments have searched for and excluded many proposed new particles
up to masses close to 1 TeV, there are many scenarios that are difficult to address at a
hadron collider.  This talk will review a number of these scenarios and present the
expectations for searches at an electron-positron collider such as the International Linear
Collider.   The cases discussed include the light Higgsino, the \stau~ slepton in the coannihilation
region relevant to dark matter, as well as other BSM signatures.
The studies are based on the ILD concept at the ILC.

}

\titlecomment{%
    *** Particles and Nuclei International Conference - PANIC2021 ***\\
  *** 5 - 10 September, 2021 ***\\
  *** Online ***

}

\def\leqsim{\mathbin{\;\raise1pt\hbox{$<$}\kern-8pt\lower3pt\hbox{$\sim$}\;}}
\def\geqsim{\mathbin{\;\raise1pt\hbox{$>$}\kern-8pt\lower3pt\hbox{$\sim$}\;}}

\def\XPM#1{\mbox{$ \tilde{\chi}^{\pm}_#1                                $}}

\def\XN#1{\mbox{$ \tilde{\chi}^0_#1                                     $}}

\def\p#1{\mbox{$ \mbox{\bf p}_1                                         $}}

\newcommand{\stau}    {\mbox{$ \tilde{\tau}                                $}}

\newcommand{\eeto}    {\mbox{$ {\, \mathrm e}^+ {\mathrm e}^- \to             $}}

\newcommand{\GeV}     {\mbox{$ {\mathrm{GeV}}                              $}}

\newcommand{\TeV}     {\mbox{$ {\mathrm{TeV}}                              $}}

\newcommand{\ba}{\begin{array}}
\newcommand{\ea}{\end{array}}
\newcommand{\bc}{\begin{center}}
\newcommand{\ec}{\end{center}}
\newcommand{\be}{\begin{eqnarray}}
\newcommand{\eeq}{\end{eqnarray}}
\newcommand{\bes}{\begin{eqnarray*}}
\newcommand{\ees}{\end{eqnarray*}}
\newcommand{\Kz}{\ifmmode {\rm K^0_s} \else ${\rm K^0_s} $ \fi}
\newcommand{\Zz}{\ifmmode {\rm Z^0} \else ${\rm Z^0 } $ \fi}
\newcommand{\xxbar}{\ifmmode {\rm x\bar{x}} \else ${\rm x\bar{x}} $ \fi}
\newcommand{\rphi}{\ifmmode {\rm R\phi} \else ${\rm R\phi} $ \fi}

\def    \missEt      {\ifmmode{/\mkern-11mu E_t}\else{${/\mkern-11mu E_t}$}\fi}
\def    \missE       {\ifmmode{/\mkern-11mu E}\else{${/\mkern-11mu E}$}\fi}
\def    \missp       {\ifmmode{/\mkern-11mu p}\else{${/\mkern-11mu p}$}\fi}
\def    \misspt      {\ifmmode{/\mkern-11mu p_t}\else{${/\mkern-11mu p_t}$}\fi}

\begin{document}
\titlepage

\section{ILC and ILD and their strong points for searches}

\begin{figure}[h] 
  \begin{center}
  \includegraphics [scale=0.3]{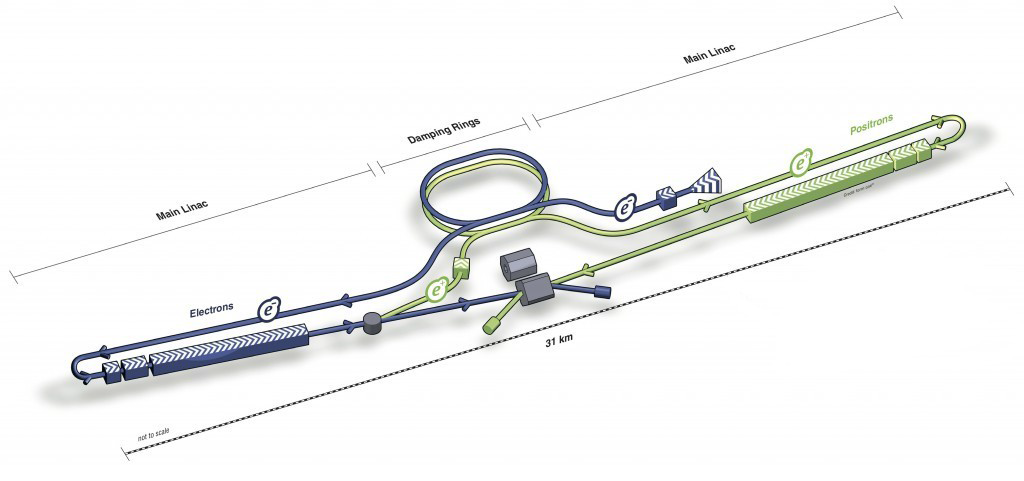}
  \includegraphics [scale=0.3]{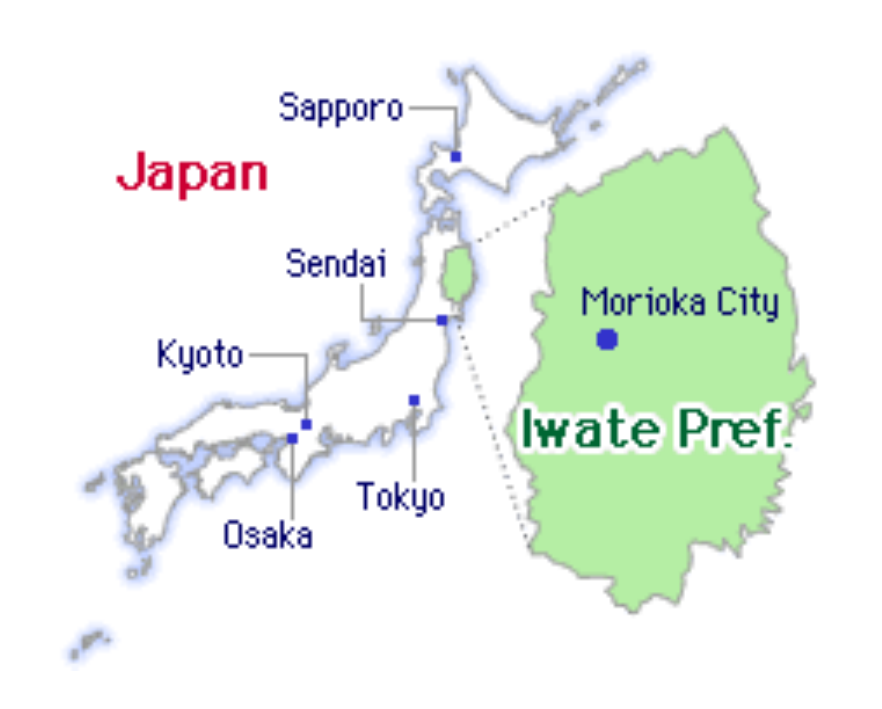}
   \end{center}
 \caption{Schematic of the ILC and the location of the proposed site in Japan's Tohoku region.\label{fig:ilc}}
 \end{figure}

The International Linear Collider (the ILC \cite{Adolphsen:2013kya}, Fig.~\ref{fig:ilc}) will collide polarised electrons with polarised positrons.
Centre-of-mass energies will range from 250 \GeV~ to 500 \GeV. The possibilities to upgrade to 1 \TeV,
and to run at $E_{CMS}= M_Z$ are also considered.
The electroweak production implied by the $e^+e^-$ initial state leads to low background rates.
This is beneficial for the detector design and optimisation: The detectors do not need to be  radiation hard,
giving the possibility to realise a  tracking system with total thickness as low as a few percent of a radiation-length.
The detector system can feature close to $ 4\pi$ coverage,
and the low rates means that it needn't be triggered, so that \textit{all}
produced events will be recorded.
Furthermore, the initial state is fully  known at an   $e^+e^-$ machine, since point-like objects are colliding.
This will be quite important for many searches for new phenomena.
The  ILC has a defined 20 year running plan, with 
integrated luminosities of 2 and 4 ab$^{-1}$  planned at $E_{CMS}=$ 250 and 500 \GeV, respectively.
It could deliver 8  ab$^{-1}$ at the possible upgrade to 1 \TeV.  
The construction of the ILC is currently under high-level political consideration in Japan.

 Both Beyond the Standard Model (BSM) searches or measurements  as well as precision SM measurements will require
that the excellent conditions provided by the accelerator are matched by excellent
 performance of the detectors.
 Specifically,
 a jet energy resolution of  3-4\%,
 an asymptotic  momentum resolution of $\sigma(1/p_\perp) = 2 \times 10^{-5}$ \GeV $^{-1}$,
and measurement of impact-parameters better than 5 $\mu$ will be needed.
 Powerful particle identification (PID) capabilities will be an asset, and
 the detector should be hermetic, with the only gaps in acceptance being the
 unavoidable vacuum pipes bringing the beams into the detector.
 Furthermore, the system should be capable to register data trigger-less.
In the International Large Detector concept (the ILD)\cite{ILDConceptGroup:2020sfq},
 having a Time Projection Chamber (TPC) as the main tracker yields the needed low mass,
 high precision, tracker
 with PID capabilities.
 The performance is further enhanced by having silicon trackers both inside and
 outside the TPC.
 The high granularity calorimeters optimised for particle flow of ILD
 assures that the needed jet energy resolution can be obtained.
 In addition, to make it possible to avoid active cooling,
 the entire system can be operated in 
 power-pulsing mode, i.e.~with the electronics being switched off
 between bunch-trains.
 
    
  \section{BSM at ILC: SUSY}

  SUSY \cite{Wess:1974tw} is the most complete theory of BSM, and therefore
  needs particular attention. In a recent contribution to the EPS-HEP 2021
  conference \cite{theotherone},
  a more extensive discussion of SUSY at ILC is made, and we only
  summarise it here.
  
    Naturalness, the hierarchy problem, the nature of dark matter (DM),
  or the observed value of the magnetic moment of the muon, all prefer
  a light electroweak sector of SUSY.
  In addition, many models
and the global set of constraints from observation
  points to a \textit{compressed spectrum}.
  If the Lightest SUSY Particle (the LSP) is Higgsino or Wino, there must be other
  bosinos close in mass to the LSP, since the $\tilde{H}$ and $\tilde{W}$
  fields have several components, leading to a close relation between
  the physical bosino states.
  Although the third possibility -  a Bino-LSP - has no such constraints,
  an overabundance of DM  is expected in this case \cite{Roy:2007ay}.
  To avoid such a situation,
  a balance between early universe LSP production and
    decay is needed.
  One compelling option is $\stau$~co-annihilation, and 
  for this process to contribute enough, the early universe density of $\stau$ and $\XN{1}$ should be similar,
  which implies that their mass must be quite similar.
  In the case of such compressed, low $\Delta(M)$, spectra, most sparticle-decays are
  via cascades,
  where the last decay in the cascade - that to SM particles and the LSP -
  features small $\Delta(M)$.
  For such decays, current LHC limits are for specific models,
  and only the limits from LEPII are model-independent.
In fact, current observations  from LHC13, LEP, g-2, DM (assumed to be 100\%~LSP),
and precision observables
taken together also point to a compressed spectrum \cite{Bagnaschi:2017tru}.
  \begin{figure}[b]
    \begin{center}
     \subcaptionbox{}{\includegraphics [align=t,scale=0.28]{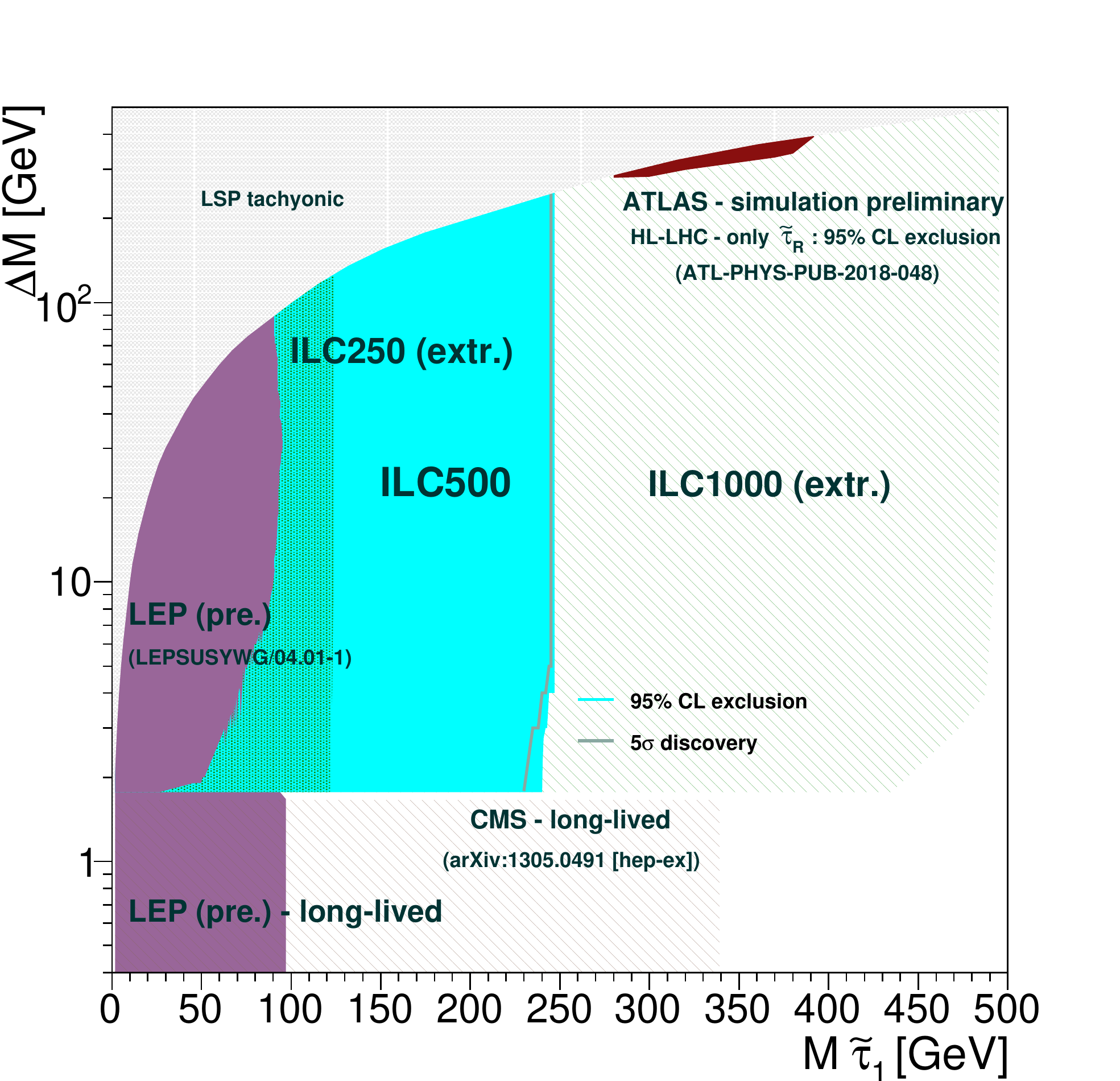}}
\subcaptionbox{}{\includegraphics [align=t,scale=0.30]{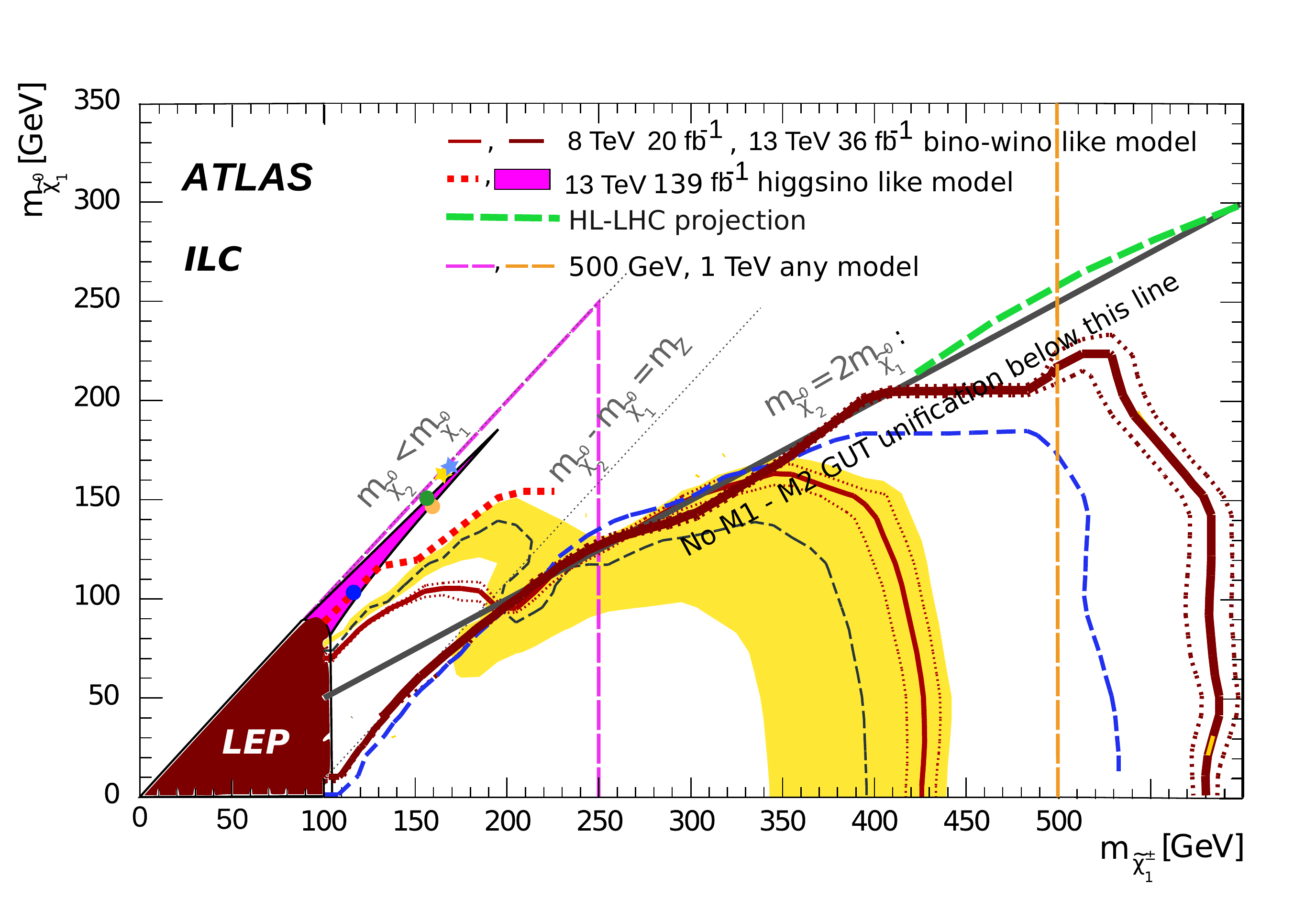}}
\end{center}
\caption{ Observed or projected exclusion regions for a $\stau$ (a) or a $\XPM{1}$ (b) NLSP, for LEPII, LHC, HL-LHC and for ILC-500 and ILD-1000\label{fig:X1summary}
}
 \end{figure}
 
 In \cite{theotherone},
 we pointed out that
at ILC, one can perform a loophole free search for SUSY
because in SUSY, the properties of NLSP
production and decay are completely predicted
for given LSP and NLSP masses.
  All possible NLSP candidates can therefore conclusively be searched for.
In Fig. \ref{fig:X1summary}, the current or projected limits are
shown, for a $\stau$ NLSP (a) \cite{NunezPardodeVera:2021cdw}, or a $\XPM{1}$ one (b) \cite{PardodeVera:2020zlr,ATLAS:2018ojr}.
In  \cite{theotherone}, it was also pointed out that these figures shows that, contrary to the LHC case,
the exclusion and discovery regions are close to identical.
From this follows that, if SUSY would be discovered at the ILC,
high precision  measurements will also be possible.
Several examples of bench-marks were shown in \cite{theotherone},
  and in all the illustrated cases, it was found that the
 SUSY masses could be determined at the sub-percent level,
 the polarised production cross-sections to the level of a few percent.
 Many other properties could also be obtained from the same data, such as
 decay branching fractions, mixing angles, and sparticle spin.



\section{BSM at ILC: New scalars, small deviations from the SM, dark photons}

\begin{figure}[b]
  \begin{center}
    \subcaptionbox{}{\includegraphics [align=c,scale=0.26]{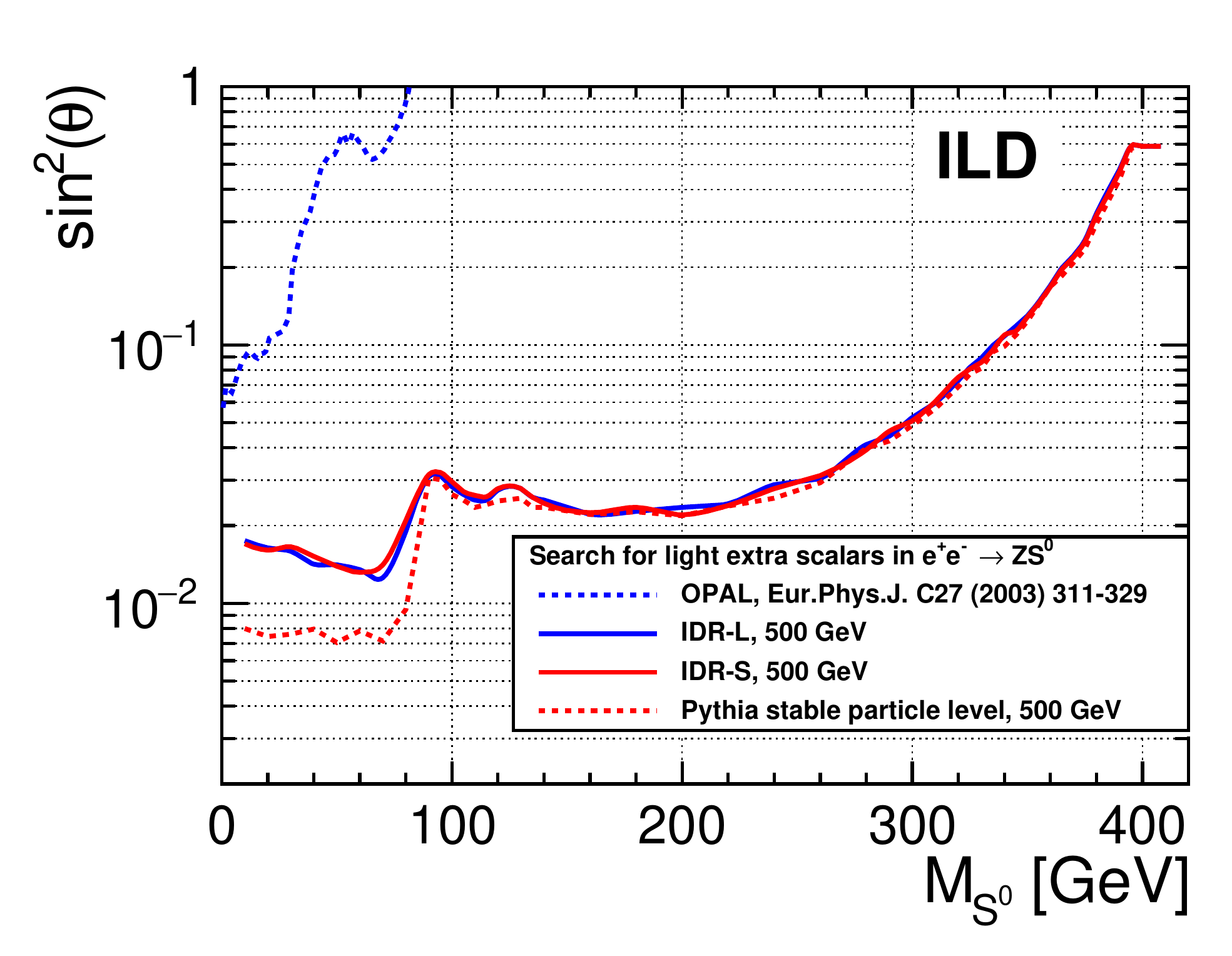}}
    \subcaptionbox{}{\includegraphics [align=c,scale=0.28]{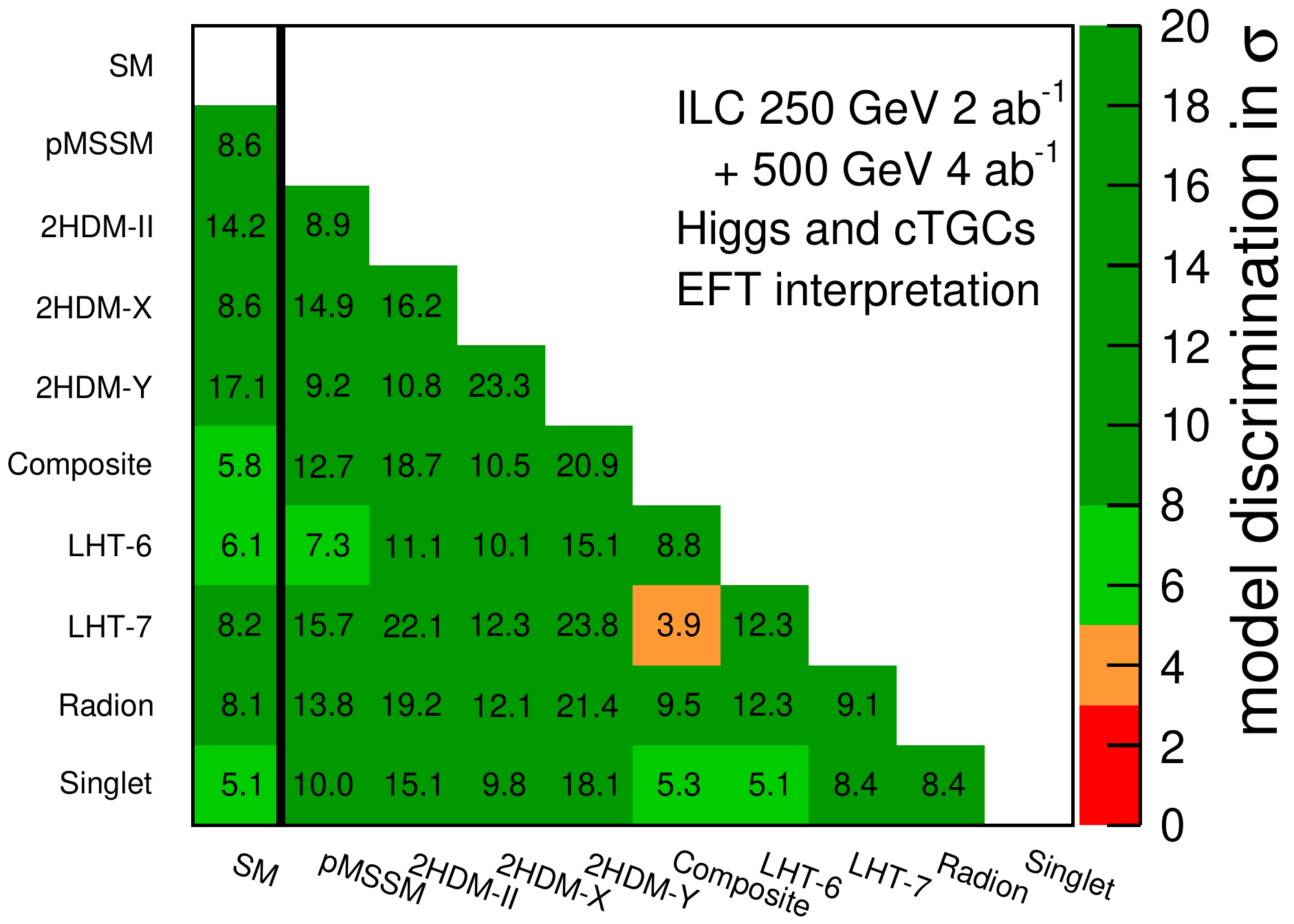}}
    \subcaptionbox{}{\includegraphics [align=c,scale=0.22]{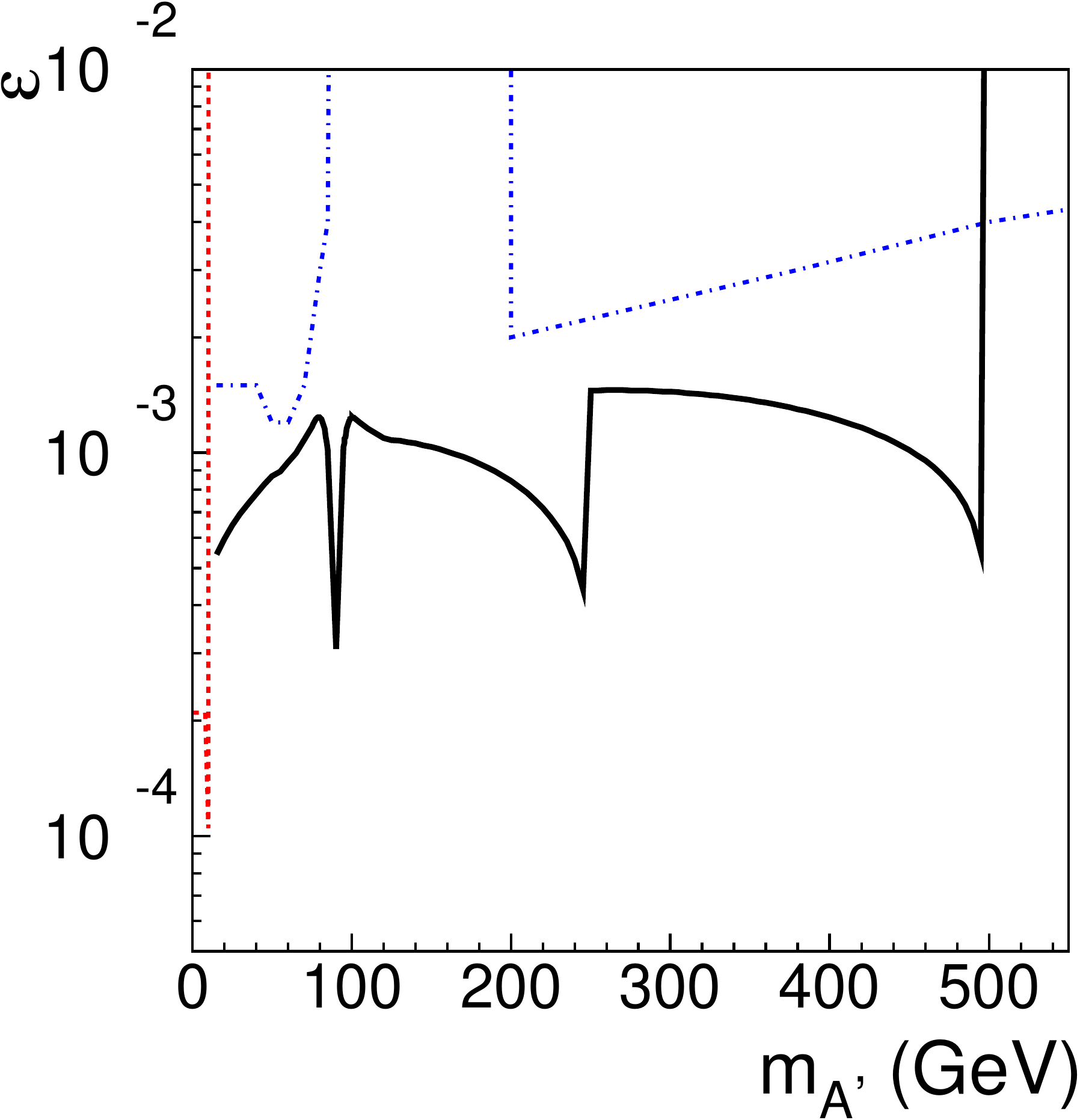}}
\end{center}
\caption{
  (a) Projected exclusion limit for
  new scalars, in terms of the coupling compared to the coupling an SM Higgs at the same mass would have.
  (b) Significances of SMEFT deviations from the expectation, both for the SM expectation and the expectation of
  each of the various listed models. (c) Exclusion limit projections for dark photons, for ILC (solid), BelleII (dash) and
  HL-LHC (dot-dash).\label{fig:otherbsm}}
 \end{figure}

Many BSM models predict the existence of a new Higgs-like scalar ($S$), produced in $\eeto Z^* $ $\rightarrow Z S$ with unknown decays of $S$.
Such a state could have escaped detection at LEP if its production cross-section is much lower than that of a SM Higgs at the same mass.
Hence, a search for such a state should be done at all accessible masses, and without any assumption on the decay modes.
At an $e^+e^-$ collider this can be done using the recoil-mass, i.e. the mass of the system recoiling against the measured $Z$.
In \cite{Wang:2020lkq}, a full ILD detector simulation study was performed, and it was found
that couplings down to a few percent of the  SM-Higgs equivalent can be excluded,  see Fig. \ref{fig:otherbsm}(a).

The ILC also offers powerful BSM discovery opportunities from indirect searches,
i.e. from detecting deviations from the  behaviour predicted by the SM.
Not only can such deviations be found, but they can also often be utilised
for model separation. 
As an example of this route to BSM physics, in Fig. \ref{fig:otherbsm}(b) we show a
Standard Model Effective Field Theory (SMEFT) study  \cite{Barklow:2017suo} using ILC results on Higgs properties and
triple gauge couplings (TGCs).
Here, the authors have selected models that are not discoverable at HL-LHC.
One observes that at the ILC, one can  both separate all the models from the SM (at the 5 $\sigma$ level),
and also separate them from each other, at a similar level on confidence.

Another BSM model that has recently received quite some attention is the
      Dark photon, the A'.
      In these models, the existence of a dark sector is postulated. This dark sector is assumed to contain
      a U(1) group, leading to the existence of a photon-like particle. This state - as well as other dark states - is assumed to be
      neutral under the SM gauge groups. However, it is likely that there would be
      kinetic mixing between ``our'' U(1) and the dark one, leading to a term $-\frac{\epsilon}{2 \cos{\theta_W}} F^\prime_{\mu\nu}B^{\mu\nu}$
      in the Lagrangian.
      Depending on the value of the free parameter $\epsilon$,
      this interaction can lead to a
    tiny, narrow resonance, which is  still
      wide enough to make decays prompt.
      Hence, the experimental approach would be to search for a narrow $\mu\mu$ resonance
       in $\eeto Z' + \mathrm{ISR} \rightarrow \mu^+ \mu^- + \mathrm{ISR}$.
     Here the excellent momentum resolution of ILD is a key issue: the better the resolution is, the
     more narrow the search-window can be, and hence the lower the background will be.
     This is also the reason why the $\mu\mu$-channel is the
      most promising one.
      In Fig. \ref{fig:otherbsm}(c), a study of this process is shown \cite{EuropeanStrategyforParticlePhysicsPreparatoryGroup:2019qin}.
      It is a theory study, but nevertheless with a quite reasonable assumption on the $M_{\mu\mu}$ resolution,
      and its dependence on the mass of the dark photon.

\section{Conclusions}
The potential for direct discovery of new particles at ILC can exceed those
of the LHC in certain, well motivated, scenarios.
This is because ILC provides
clean environment without QCD backgrounds, and a  well-defined initial state.
Furthermore, detectors at the ILC, such as ILD, will be more precise, will be hermetic, and will not need
to be triggered.
In addition, ILC also is extendable in energy and features polarised beams. 

Synergies between ILC and LHC are expected:  experiments at LHC will have higher energy-reach,
while those at ILC
will be more sensitive for subtle signals.
For instance, if SUSY is reachable at the ILC, precision measurements can be done.
This input would help in the interpretation of anomalies seen at the LHC, and
might even be what is needed to transform a 3 $\sigma$ excess to a discovery of states
  beyond the reach of ILC.



\section{Acknowledgements}
We would like to thank the LCC generator working group and the ILD software
working group for providing the simulation and reconstruction tools and
producing the Monte Carlo samples used in this study.
This work has benefited from computing services provided by the ILC Virtual
Organisation, supported by the national resource providers of the EGI
Federation and the Open Science GRID.

\end{document}